\documentclass[aps,prl,twocolumn,showpacs,superscriptaddress,groupedaddress,preprintnumbers,floatfix,reprint,superscriptaddress]{revtex4-1}

\usepackage{graphicx}  
\usepackage{dcolumn}   
\usepackage{bm}        
\usepackage{amssymb}   

\usepackage{subfigure}
\usepackage{multirow}
\usepackage{placeins}
\usepackage{morefloats}
\usepackage{epstopdf}

\usepackage{amsfonts}    
\usepackage{amsmath}
\usepackage{mathtools}

\newcommand{\Rsd}{\prescript{\ell}{}{R^s_d}}

\begin{document}

\preprint{
\vbox{
\hbox{ADP-14-10/T868}
\hbox{Edinburgh 2014/08}
\hbox{LTH 1005}
\hbox{DESY 14-036}
}}

\title{Determination of the strange nucleon form factors}

\author{P.E.~Shanahan}\affiliation{ARC Centre of Excellence in Particle Physics at the Terascale and CSSM, Department of Physics, University of Adelaide, Adelaide SA 5005, Australia}
\author{R.~Horsley}\affiliation{School of Physics and Astronomy, University of Edinburgh, Edinburgh EH9 3JZ, UK}
\author{Y.~Nakamura}\affiliation{RIKEN Advanced Institute for Computational Science, Kobe, Hyogo 650-0047, Japan}
\author{D.~Pleiter}\affiliation{JSC, Forschungzentrum J\"ulich, 52425 J\"ulich, Germany} \affiliation{Institut f\"ur Theoretische Physik, Universit\"at Regensburg, 93040 Regensburg, Germany}
\author{P.E.L.~Rakow}\affiliation{Theoretical Physics Division, Department of Mathematical Sciences, University of Liverpool, Liverpool L69 3BX, UK}
\author{G.~Schierholz}\affiliation{Deutsches Elektronen-Synchrotron DESY, 22603 Hamburg, Germany}
\author{H.~St\"uben}\affiliation{Regionales Rechenzentrum, Universit\"at Hamburg, 20146 Hamburg, Germany}
\author{A.W.~Thomas}\affiliation{ARC Centre of Excellence in Particle Physics at the Terascale and CSSM, Department of Physics, University of Adelaide, Adelaide SA 5005, Australia}
\author{R.D.~Young}\affiliation{ARC Centre of Excellence in Particle Physics at the Terascale and CSSM, Department of Physics, University of Adelaide, Adelaide SA 5005, Australia}
\author{J.M.~Zanotti}\affiliation{ARC Centre of Excellence in Particle Physics at the Terascale and CSSM, Department of Physics, University of Adelaide, Adelaide SA 5005, Australia}

\collaboration{CSSM and QCDSF/UKQCD Collaborations}

\begin{abstract}
The strange contribution to the electric and magnetic form factors of the nucleon is determined at a range of discrete values of $Q^2$ up to $1.4$~GeV$^2$. This is done by combining a recent analysis of lattice QCD results for the electromagnetic form factors of the octet baryons with experimental determinations of those quantities. The most precise result is a small negative value for the strange magnetic moment: $G_M^s(Q^2=0) = -0.07\pm0.03\,\mu_N$. At larger values of $Q^2$ both the electric and magnetic form factors are consistent with zero to within two standard deviations.
\end{abstract}


\maketitle

A quantitative determination of the contribution of non-valence flavour quarks to nucleon observables remains a fundamental challenge of hadronic physics. Since such contributions must arise entirely through interactions with the vacuum, their sign and magnitude provide key information regarding the nonperturbative structure of the nucleon; their determination within non-perturbative QCD constitutes a test of a level of importance comparable to that of the Lamb shift for QED. Strange quarks, as the lightest sea-only flavour, are expected to play the largest role.

Recent years have seen extensive experimental efforts directed at measuring strangeness in the nucleon. The strange electromagnetic form factors in particular have been determined from experiments at JLab (G0, HAPPEX)~\cite{Armstrong:2005hs,Androic:2009aa,Aniol:2005zg,Aniol:2005zf,Acha:2006my,PhysRevLett.108.102001,PhysRevC.69.065501}, MIT-Bates (SAMPLE)~\cite{Spayde200479,Beise:2004py}, and  Mainz (A4)~\cite{Maas:2004dh,Maas:2004ta,Baunack:2009gy}. Probing a range of values of $Q^2$ up to $\approx 0.94~$GeV$^2$, the combined data sets constrain the strange contribution to the nucleon form factors to be less than a few percent but are consistent with zero to within 2-sigma~\cite{Young:2006jc}. 
The status of the strange form factors from theory is less clear; predictions from various quark models cover a very broad range of values~\cite{PhysRevLett.95.072001,Jaffe1989275,Hammer1996323,Cohen19931,PhysRevD.43.869,Weigel199520}, and the large computational cost of all-to-all propagators has so far limited direct lattice QCD studies to large pion masses and single volumes~\cite{Abdel-Rehim:2013wlz,Doi:2009sq}.

In this Letter we determine the strangeness contributions to the nucleon electromagnetic form factors indirectly at a range of values of $Q^2$ currently unattainable through direct experimental measurement. 
Under the assumption of charge symmetry, one can combine experimental measurements of the total nucleon form factors with lattice QCD determinations of the connected (or `valence' quark) contributions to deduce the disconnected (or `sea' quark) components~\cite{Leinweber:1995ie}. 
This method has been applied previously to determine the strange magnetic form factor at $Q^2=\{0,0.23\}$~GeV$^2$~\cite{Leinweber:2004tc,Wang:1900ta} and the strange electric form factor at $Q^2=0.1$~GeV$^2$~\cite{Leinweber:2006ug} from quenched lattice QCD results. 
In this work we are able to perform a complete study using a recent analysis of dynamical $2+1$--flavour lattice QCD simulations~\cite{Shanahan:2014Elec,Shanahan:2014uka} to determine both the strange electric and magnetic form factors at six discrete values of $Q^2$ up to 1.4~GeV$^2$.

The lattice results used here are an extension of those reported in Refs.~\cite{Shanahan:2014Elec,Shanahan:2014uka}; we include two independent sets of $2+1$-flavour simulations at different values of the finite lattice spacing $a$. The lattice volumes are $L^3\times T=32^3 \times 64$ and $48^3 \times 96$, and the lattice spacings are $a=0.074(2)$~fm and $0.062(2)$~fm (set using various singlet quantities~\cite{Horsley:2013wqa,Bietenholz:2011qq}) for the two sets respectively. 
The particular values used as input here are the connected quark contributions to the electric and magnetic form factors of the outer-ring octet baryons after extrapolation to infinite volume and to the physical pseudoscalar masses. That extrapolation, detailed in Refs.~\cite{Shanahan:2014Elec,Shanahan:2014uka}, is performed using a formalism based on connected chiral perturbation theory~\cite{Leinweber:2002qb,Tiburzi:2009yd}.

The extraction of the strange electromagnetic form factors from the extrapolated lattice results follows the procedure introduced in Refs.~\cite{Leinweber:1998kz,Leinweber:1999nf}.
Under the assumption of charge symmetry, which is an exact symmetry of QCD if one neglects QED and the light quark mass difference (i.e., assuming $m_u=m_d$), one may express the electromagnetic form factors of the proton and neutron as~\cite{Leinweber:1995ie}
\begin{align}
\label{eq:p}
p & = e^uu^p + e^dd^p + O_N, \\ \label{eq:n}
n & = e^du^p + e^ud^p + O_N.
\end{align}
Here, $p$ and $n$ denote the physical (electric or magnetic) form factors of the proton and neutron and $u^p$ and $d^p$ represent the connected $u$ and $d$ quark contributions to the proton form factor. 
The disconnected quark loop term, $O_N$, may be decomposed into individual quark contributions:
\begin{align}
O_N & =\frac{2}{3}\prescript{\ell}{}{G}^u-\frac{1}{3}\prescript{\ell}{}{G}^d-\frac{1}{3}\prescript{\ell}{}{G}^s, \\
& = \frac{\prescript{\ell}{}{G}^s}{3}\left(\frac{1-\Rsd}{\Rsd}\right),
\label{eq:On}
\end{align}
where charge symmetry has been used to equate $\prescript{\ell}{}{G}^u=\prescript{\ell}{}{G}^d$ and the ratio of $s$ to $d$ disconnected quark loops is denoted by $\Rsd = \prescript{\ell}{}{G}^s/\prescript{\ell}{}{G}^d$.

Rearranging Eqs.~(\ref{eq:p}), (\ref{eq:n}) and (\ref{eq:On}) to isolate the strange quark loop contribution $\prescript{\ell}{}{G}^s$ yields two independent expressions which are rigorous consequences of QCD under the assumption of charge symmetry:
\begin{align}
\label{eq:ls01}
\prescript{\ell}{}{G}^s & = \left( \frac{\Rsd}{1-\Rsd}\right)\left[2p+n-u^p\right], \\ \label{eq:ls02}
\prescript{\ell}{}{G}^s & = \left( \frac{\Rsd}{1-\Rsd}\right)\left[p+2n-d^p\right].
\end{align}
In principle, given a suitable estimate of $\Rsd$, these expressions may be simply evaluated; the total form factors $p$ and $n$ are well known experimentally and the connected contributions $u^p$ and $d^p$ may be calculated on the lattice.

This procedure relies on the assumption that the difference between the experimental numbers and the connected lattice simulation results for the form factors may be entirely attributed to contributions from disconnected quark loops, i.e., that all other systematic effects are under control. 
In order to be able to estimate any as-yet undetermined lattice systematics, we average Eqs.~(\ref{eq:ls01}) and (\ref{eq:ls02}) resulting in a form where only the connected contribution to the combination $(u^p+d^p)_\text{conn.}$ needs to be determined from the lattice: 
\begin{align}
\label{eq:newEll}
\prescript{\ell}{}{G}^s & = \left(\frac{\Rsd}{1-\Rsd}\right)\left[ \frac{3}{2}(p+n)-\frac{1}{2}(u^p+d^p)_\textrm{conn.}\right].
\end{align}
Relaxing the assumption of exact charge symmetry in the valence sector would result in an additional term $+\frac{3}{2}G^{u,d}$ (where, in the notation of Ref.~\cite{PhysRevC.74.015204}, $G^{u,d}$ is the systematic CSV uncertainty affecting experimental determinations of the strange form factors) appearing within the square brackets of Eq.~\eqref{eq:newEll}. For low values of $Q^2$ in particular, where $\left(\Rsd/(1-\Rsd)\right)$ is small, this systematic thus affects our extraction of the strange form factors considerably less than it impacts on experimental determinations of these quantities, where the assumption of good charge symmetry is also standard. 
Taking the values of $G^{u,d}$ from Ref.~\cite{PhysRevC.74.015204} as a systematic uncertainty would increase our error bands by less than 10\%.
%
%
Furthermore, a recent re-evaluation of $G^{u,d}$ using relativistic chiral perturbation theory with a more realistic $\omega$-nucleon coupling~\cite{Wagman:2014nfa} found a significant reduction in $G^{u,d}$, suggesting that the assumption of good charge symmetry has a negligible effect on our results.   
For values of $Q^2$ larger than about 0.3~GeV$^2$ there have been few calculations of the relevant CSV quantities to date. However, a lattice-based determination using the same simulations used for this work, independent of assumptions regarding strangeness, suggests that CSV effects remain negligible for this calculation of the strange form factors across the entire $Q^2$-range of relevance~\cite{ShanahanCSV}.

We discuss in turn each of the three inputs into Eq.~(\ref{eq:newEll}):
\begin{itemize}
\item The lattice values for $(u^p+d^p)_\text{conn.}$.
\item The experimental $p$ and $n$ form factors.
\item The ratio $\Rsd={^\ell} G^s/{^\ell G^d}$.
\end{itemize}
As described previously, the lattice results used for the connected $u$ and $d$ quark contributions to the proton electric and magnetic form factors, $u^p$ and $d^p$, are an extended set of those presented in Refs.~\cite{Shanahan:2014Elec,Shanahan:2014uka}. Both statistical uncertainties and systematic effects resulting from the chiral and infinite-volume extrapolations, including an estimate of the model-dependence, are accounted for.
We additionally allow for any unknown systematics on the combination $(u^p+d^p)_\text{conn.}$ by estimating that such effects will be similar in magnitude for the isovector combination $(u^p-d^p)_\text{conn.}$ which may be directly compared with experiment. 
Because disconnected contributions in the total form factors cancel in the combination $(p-n)$, the difference $(u^p-d^p)_\textrm{Latt.}-(p-n)_\textrm{Exp.}$ provides an estimate of any unaccounted-for uncertainty in the lattice simulation results. 
We take the largest value of this difference, evaluated at range of discrete simulation values of $Q^2$, as a conservative estimate.

This procedure is followed for both the electric and magnetic form factors. The additional uncertainty included in this fashion is significant and larger than the statistical uncertainty in the determination of the strange magnetic form factor. For the electric form factor it is a modest contribution of a size similar to or smaller than the statistical uncertainty.

The total proton and neutron electromagnetic form factors $p$ and $n$ are taken from the parameterizations of experimental results by Kelly~\cite{Kelly:2004hm} and Arrington and Sick~\cite{Arrington:2006hm} (the latter is used only on its quoted range of validity, $Q^2 <1$~GeV$^2$). The entire calculation, including the additional estimate of lattice systematics, is performed using each parameterization. 
The average central value of the two sets of results is taken as the best-estimate of the strange form factors.
Half of the difference between the two central values is included as an estimate of the parameterization-dependent uncertainty. 
This contribution to the uncertainty is small.

We derive an estimate for the disconnected quark-loop ratio $\Rsd={^\ell G^s}/{^\ell G^d}$ using a model based on chiral effective field theory, as also done in Refs.~\cite{Leinweber:2004tc,Wang:1900ta,Leinweber:2006ug}. In that formalism $\Rsd$ is given by the ratio of loop diagram contributions to the electromagnetic form factors, where the relevant loop integrals are weighted by the appropriate `disconnected' chiral coefficients for the $s$ and $d$ quarks~\cite{Leinweber:2002qb,Wang:1900ta,Leinweber:2006ug}.  

\begin{figure}
\begin{center}
\subfigure[]{\label{fig:mesinsoct}\includegraphics[width=0.235\textwidth]{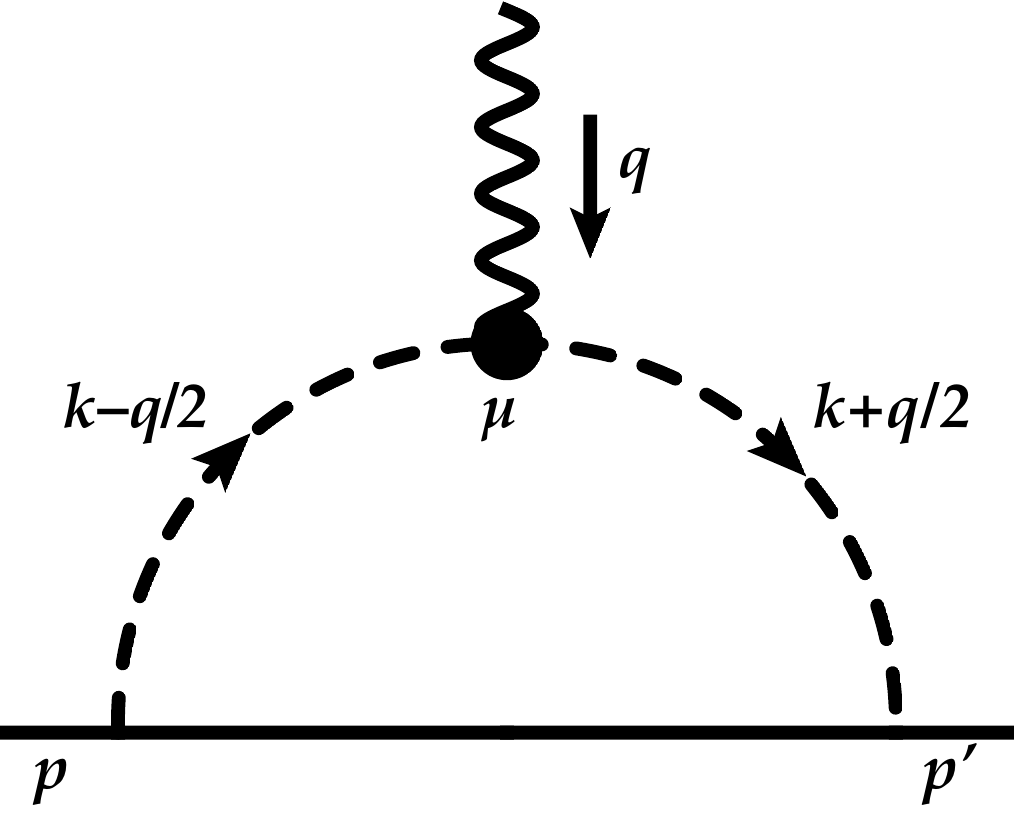}}
\subfigure[]{\label{fig:octinsoct}\includegraphics[width=0.235\textwidth]{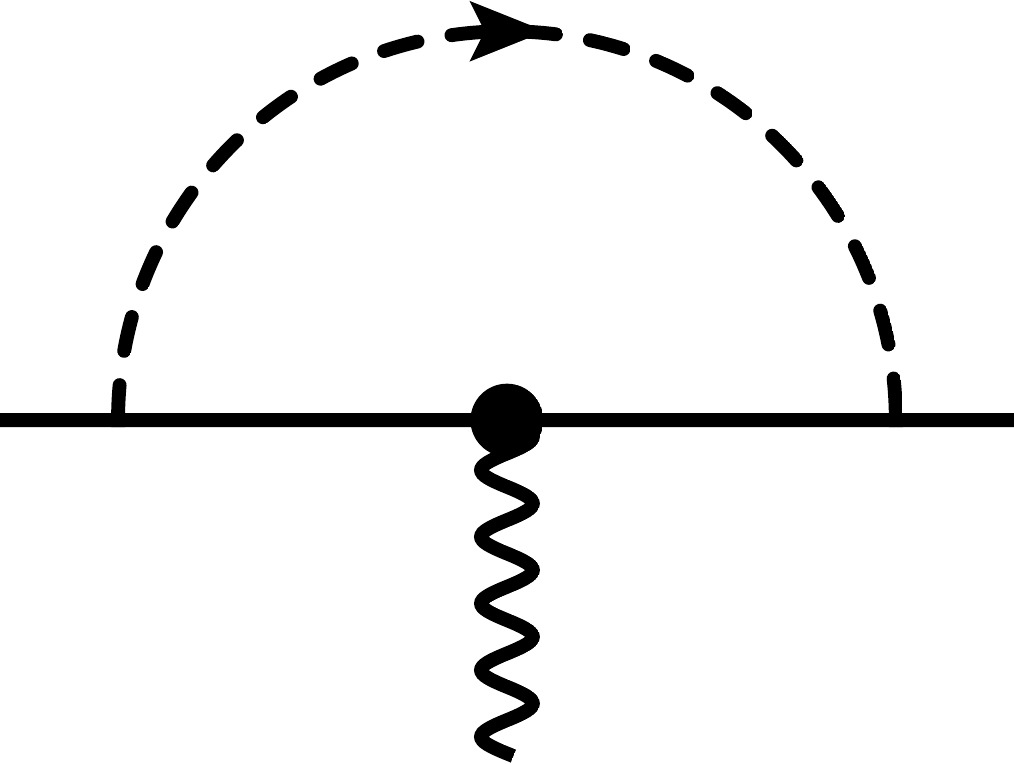}}
\caption{
Loop diagrams which are included in the estimate of $\Rsd$ from effective field theory. Fig.~\ref{fig:octinsoct} is included for the electric form factor only. The solid, dashed and wavy lines denote octet baryons, mesons and photons respectively.}
\label{fig:mesinsloopst}
\end{center}
\end{figure}

The primary loop diagram relevant to this calculation is depicted in Fig.~\ref{fig:mesinsoct}. For the electric form factor in particular, a higher-order diagram (Fig.~\ref{fig:octinsoct}) is important as it makes a significant contribution of the opposite sign to that of Fig.~\ref{fig:mesinsoct}, resulting in a large cancellation.
While to the order of the calculation in Refs.~\cite{Shanahan:2014Elec,Shanahan:2014uka} this term contributes a constant to $G_E(Q^2)$ (enforcing charge conservation at $Q^2=0$), this is not a good approximation for the large $Q^2$ values considered in this work. 

For this reason we include Fig.~\ref{fig:octinsoct}, with an estimate of its $Q^2$-dependence, explicitly in our calculation of $\Rsd$ for the electric form factor. This is achieved by calculating the diagram in heavy-baryon chiral perturbation theory and modelling the $Q^2$-dependence of the photon-baryon vertex based on the lattice results of Ref.~\cite{Shanahan:2014Elec}.

\begin{figure}
\begin{center}
\includegraphics[width=0.48\textwidth]{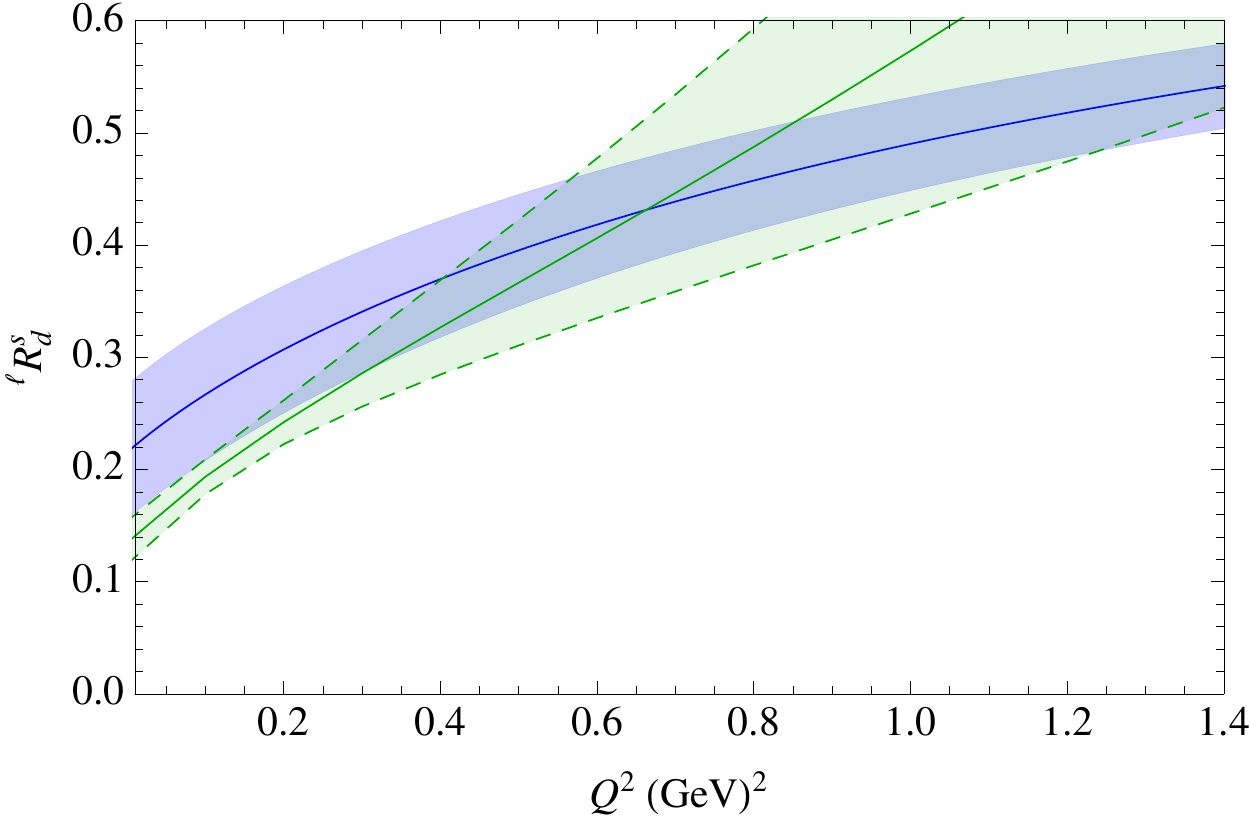}
\caption{Estimate of $\Rsd$ from effective field theory with finite-range regularization for the electric (dashed green) and magnetic (solid blue) form factors.}
\label{fig:Rsd}
\end{center}
\end{figure}

The uncertainty in the ratio $\Rsd$ is estimated by additionally including loops with decuplet-baryon intermediate states, as well as allowing the dipole mass parameter $\Lambda$ used in the finite-range regularization scheme to vary between 0.6 and 1.0~GeV~\cite{Leinweber:2003dg,Young:2002cj,Young:2002ib}. The resulting values for $\Rsd$ are shown in Fig.~\ref{fig:Rsd}. 

\begin{figure}
\begin{center}
\subfigure[]{\label{fig:GMs}\includegraphics[width=0.48\textwidth]{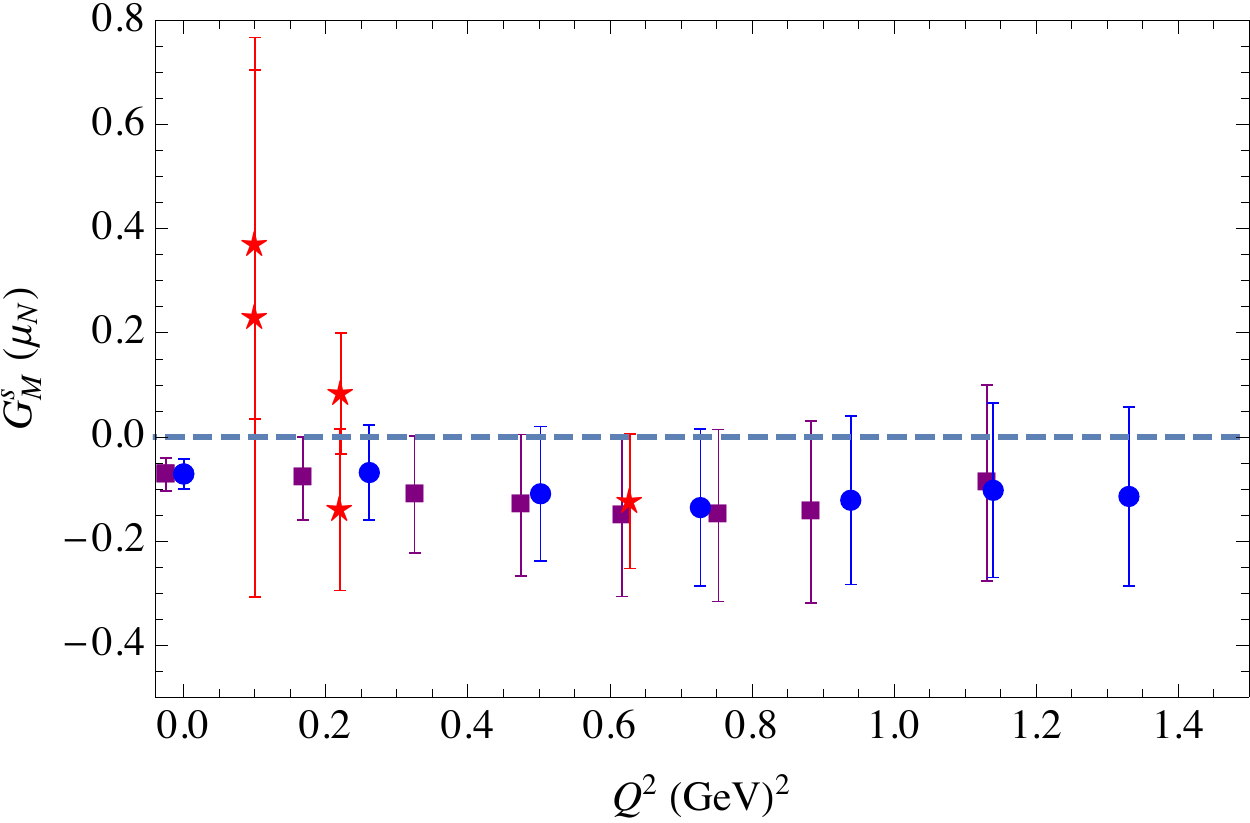}}
\subfigure[]{\label{fig:GEs}\includegraphics[width=0.48\textwidth]{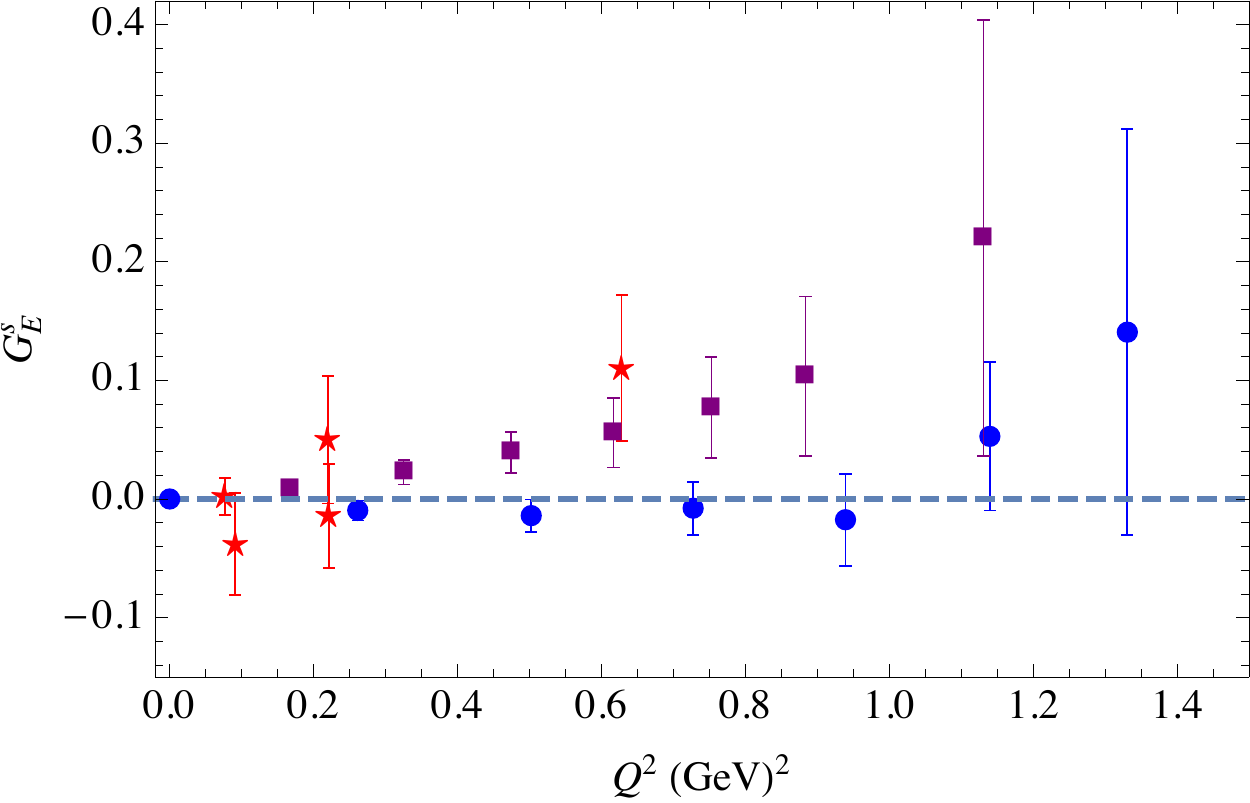}}
\caption{Strange contribution to the magnetic (\ref{fig:GMs}) and electric (\ref{fig:GEs}) form factors of the proton, for strange quarks of unit charge. The blue circles and purple squares show the results of independent analyses based on lattice simulations with scales $a=0.074(2)$~fm and $0.062(2)$~fm respectively. The experimental results (red stars) are taken from Refs.~\cite{Acha:2006my,Aniol:2005zf,Androic:2009aa,Baunack:2009gy,Beise:2004py,Spayde200479}.}
\label{fig:StrangeFFs}
\end{center}
\end{figure}

Finally, the results of this analysis (using Eq.\eqref{eq:newEll}) for the strange electric and magnetic form factors of the proton at non-zero $Q^2$ are summarized in Table~\ref{tab:results} and are displayed in Fig.~\ref{fig:StrangeFFs} alongside the latest experimental determinations of those quantities.
All results (away from $Q^2=0$) are consistent with zero to within 2-sigma. The results for the strange magnetic form factor favour negative values which are consistent with recent experimental results. For the electric form factor, the two independent analyses based on lattice QCD simulations at different lattice spacings and volumes are inconsistent at 1-sigma.
As a result, simple estimates of the strange electric charge radius of the proton using a straight-line fit in $Q^2$ to the lowest-$Q^2$ result for $G_E^s$ give results with opposite signs for the two analyses: 
\begin{equation}
\langle r^2_E \rangle^s = 
\begin{cases}
0.0086(79)~\text{fm}^2, & a=0.074(2)\,\text{fm}\\
-0.0114(88)~\text{fm}^2, & a=0.062(2)\,\text{fm}
\end{cases}
\end{equation}
Although we cannot make a conclusive statement without additional simulation results, we expect that this difference is dominated by statistical fluctuations.

\begin{figure}
\begin{center}
\includegraphics[width=0.47\textwidth]{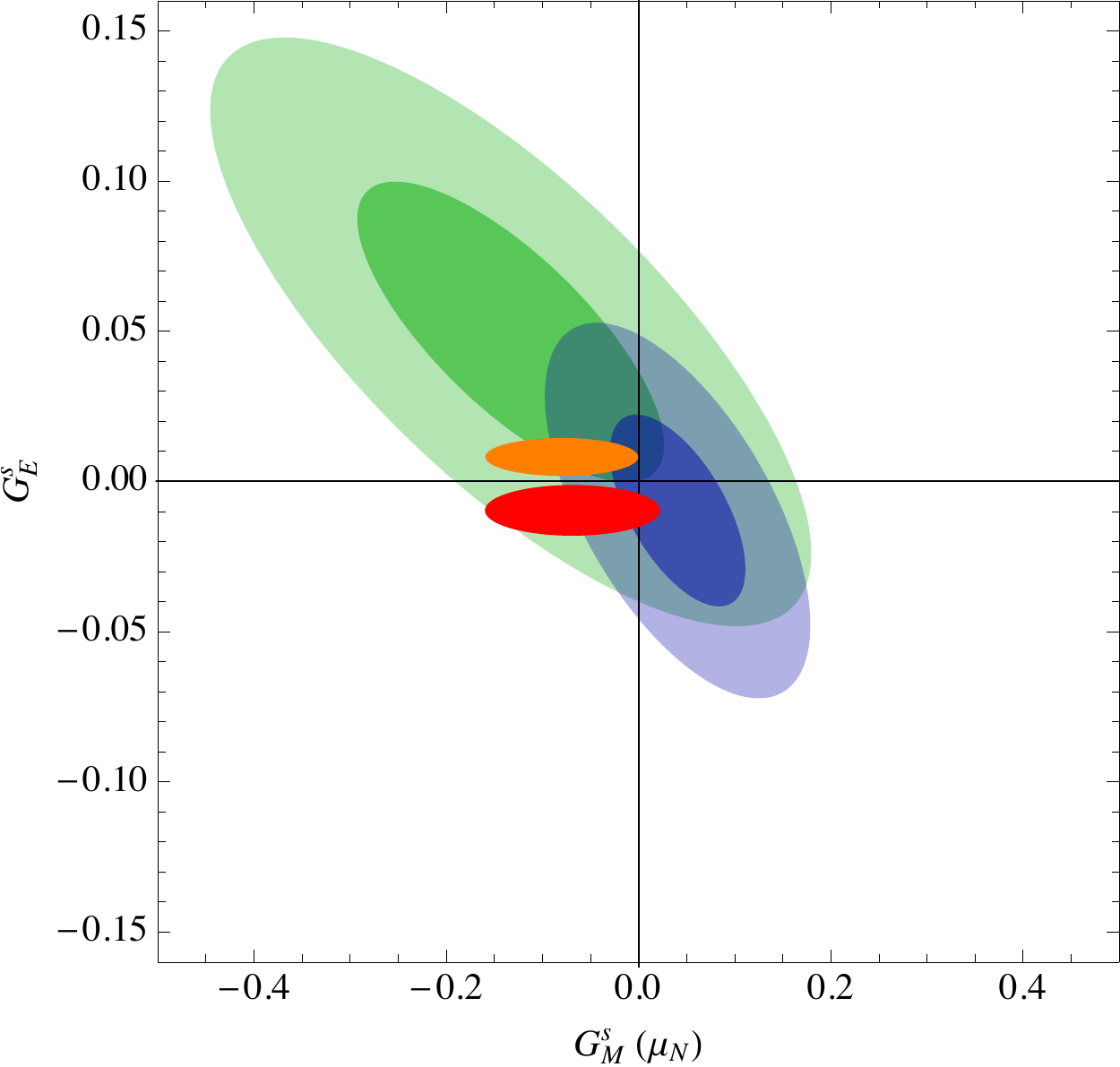}
\caption{Comparison of the results of this work to 1-sigma (red ellipse) at $Q^2=0.26$~GeV$^2$ for $a=0.074(2)$~fm and (orange ellipse) at $Q^2=0.17$~GeV$^2$ at $0.062(2)$~fm with available experimental results at similar values of $Q^2$. The dark and pale green ellipses show 1-sigma and 2-sigma results from the A4 Collaboration at $Q^2=0.23$ GeV$^2$~\cite{Maas:2004ta} while the blue ellipses show G0 Collaboration results close to $Q^2=0.23$~GeV$^2$~\cite{Armstrong:2005hs,Androic:2009aa}.}
\label{fig:StrangeFFsOp22}
\end{center}
\end{figure}

Since experimental determinations of the strange form factors are obtained as linear combinations of $G^s_E$ and $G^s_M$ we also display results at the lowest values of the momentum transfer, $Q^2=0.26$\,GeV$^2$ and 0.17\,GeV$^2$ for the $a=0.074(2)$\,fm and $0.062(2)$\,fm simulation sets respectively, in the $G^s_M$-$G^s_E$ plane in Fig.~\ref{fig:StrangeFFsOp22}. 
The available experimental results for similar values of $Q^2$ appear on this figure as ellipses.
Both present calculations are consistent with experiment to within 2-sigma.

\begin{table}
\ruledtabular
\begin{center}
\begin{tabular}{cccc}
$a$ (fm) & $Q^2$ (GeV$^2$) & $G^s_M$ ($\mu_N$) &$G^s_E$  \\\hline
0.074(2) &0.26 & -0.069(91) & -0.096(84)\\
&0.50 & -0.11(13) & -0.014(14)\\
&0.73 & -0.14(15) & -0.008(22)\\
&0.94 & -0.12(16) & -0.017(39)\\
&1.14 &  -0.10(17) & 0.053(62)\\
&1.33 &  -0.12(17) & 0.14(17) \\ \hline
0.062(2)&0.17 & -0.080(80) & 0.0081(63)\\
&0.33 & -0.11(11) & 0.023(10)\\
&0.47 & -0.13(14) & 0.039(17)\\
&0.62 & -0.15(15) & 0.056(29)\\
&0.75 & -0.15(17) & 0.077(43)\\
&0.88 & -0.14(17) & 0.104(67)\\
&1.13 & -0.089(188) & 0.22(18)
\end{tabular}
\end{center}
\caption{Results for the strange electric and magnetic form factors of the proton with all contributions to the uncertainty combined in quadrature. The two sets of results correspond to independent analyses based on lattice simulations with scales $a=0.074(2)$~fm and $0.062(2)$~fm respectively.}
\label{tab:results}
\end{table}


Using the additional information available from experiment at $Q^2=0$, where the hyperon form factors have been measured~\cite{pdg}, we also determine the strange contribution to the proton magnetic moment. 
We rearrange Eqs.~(\ref{eq:ls01}) and (\ref{eq:ls02}), using the assumption of charge symmetry, to express the nucleon strange magnetic moment in terms of the hyperon moments~\cite{Leinweber:1995ie,Leinweber:1999nf}:
\begin{align}
\label{eq:ls1}
\prescript{\ell}{}{G}^s & = \left( \frac{\Rsd}{1-\Rsd}\right)\left[2p+n - \frac{u^p}{u^\Sigma}\left(\Sigma^+-\Sigma^-\right)\right], \\ \label{eq:ls2}
\prescript{\ell}{}{G}^s & = \left( \frac{\Rsd}{1-\Rsd}\right)\left[p+2n - \frac{u^n}{u^\Xi}\left(\Xi^0-\Xi^-\right)\right].
\end{align}
This rearrangement minimizes the propagation of lattice systematics as only ratios of form factors must be determined from lattice QCD.

The ratios $u^p_M/u^\Sigma_M$ and $u^n_M/u^\Xi_M$ of connected up quark contributions to the hyperon form factors, at a range of non-zero values of the momentum transfer $Q^2$, are taken from the lattice QCD analyses described earlier~\cite{Shanahan:2014Elec,Shanahan:2014uka}. 
We determine the $Q^2=0$ values needed here using a linear extrapolation in $Q^2$, with an additional experimental constraint provided by the equality of Eqs.~(\ref{eq:ls1}) and (\ref{eq:ls2}):
\begin{equation}
\frac{u^p_M}{u^\Sigma_M}=\frac{u^n_M}{u^\Xi_M}\left(\frac{\mu_{\Xi^0}-\mu_{\Xi^-}}{\mu_{\Sigma^+}-\mu_{\Sigma^-}}\right) + \left(\frac{\mu_{p}-\mu_{n}}{\mu_{\Sigma^+}-\mu_{\Sigma^-}}\right),
\label{eq:rats}
\end{equation}
where $\mu_B$ denotes the experimental magnetic moment of the baryon $B$~\cite{pdg}. The fit is performed to the lattice results where $Q^2<1$~GeV$^2$, which display qualitatively linear behaviour and for which the linear-fit $\chi^2/$d.o.f is acceptable given the constraint of Eq.~(\ref{eq:rats}). Fitting to one less data point does not change the results to the precision quoted.

The best estimates of the $Q^2=0$ ratios of connected contributions to the baryon magnetic form factors are
\begin{equation}
\left[\frac{u^p_M}{u^\Sigma_M},\frac{u^n_M}{u^\Xi_M}\right] = 
\begin{cases} 
\left[1.096(16), 1.239(90)\right], & a=0.074(2)\,\text{fm} \\ 
\left[1.095(17), 1.222(98)\right], & a=0.062(2)\,\text{fm}
\end{cases}
\end{equation}
where the two sets of results correspond to our two independent analyses using lattice QCD simulation results at different lattice spacings and volumes as described earlier. These full-QCD numbers align remarkably well with those determined in Ref.~\cite{Leinweber:2004tc}, given that that analysis was based on quenched lattice simulation results after the application of a theoretical `unquenching' formalism~\cite{Young:2002cj}. 

The resulting values for the strange magnetic moment (from Eqs.~(\ref{eq:ls1}) and (\ref{eq:ls2})), conventionally defined without the charge factor, are
\begin{equation}
G_M^s(Q^2=0) = 
\begin{cases}
-0.071(13)(25)(4) \, \mu_N, & a=0.074(2)\,\text{fm} \\
-0.073(14)(26)(4) \, \mu_N, & a=0.062(2)\,\text{fm}
\end{cases}
\end{equation}
The first uncertainty is propagated from the lattice simulation results, the second, dominant, contribution comes from the ratio $\Rsd$ and the last is that from the experimental determination of the magnetic moments~\cite{pdg}. 
Clearly, the results of our analysis using two independent calculations performed at different lattice spacings and volumes are in excellent agreement.

Our final result for the strange magnetic moment of the proton, $G_M^s(Q^2=0) = -0.07\pm0.03 \,\mu_N$, is non-zero to 2-sigma and an order of magnitude more precise than the closest experimental results.
The results reported at the values of $Q^2$ above 0.6~GeV$^2$ are the first determinations, experimental or based on lattice QCD, in that region. At present they cannot be distinguished from zero, but the uncertainties constrain their actual values to be very small.

\section*{Acknowledgements}
We thank D.~B.~Leinweber for a careful and critical reading of the manuscript. The numerical configuration generation was performed using the BQCD lattice QCD program~\cite{Nakamura:2010qh} on the IBM BlueGeneQ using DIRAC 2 resources (EPCC, Edinburgh, UK), the BlueGene P and Q at NIC (J\"ulich, Germany) and the Cray XC30 at HLRN (Berlin-Hannover, Germany). The BlueGene codes were optimised using Bagel~\cite{Boyle:2009vp}. The Chroma software library~\cite{Edwards:2004sx} was used in the data analysis. This work was supported by the EU grants 283286 (HadronPhysics3), 227431 (Hadron Physics2) and by the University of Adelaide and the Australian Research Council through the ARC Centre of Excellence for Particle Physics at the Terascale and grants FL0992247 (AWT), FT120100821 (RDY), DP140103067 (RDY and JMZ) and FT100100005 (JMZ).

\section*{References}
\bibliography{StrangeBib}

\end{document}